\documentclass[aps,pre,twocolumn,notitlepage,prl]{revtex4-2}
\usepackage{amsmath,amssymb,bm}
\usepackage{graphicx}
\usepackage{xcolor}
\usepackage[percent]{overpic}
\usepackage{physics}
\usepackage{hyperref}
\hypersetup{hidelinks}
\usepackage{enumitem}
\usepackage{comment}

\begin{document}

\title{
When dissipative steady states admit thermodynamic occupation laws
}

\author{Tetsu Ichitsubo}\email{tichi@tohoku.ac.jp}
\affiliation{
Institute for Materials Research, Tohoku University, Sendai 980-8577, Japan\\
Advanced Science Research Center, Japan Atomic Energy Agency, Tokai, Ibaraki 319-1195, Japan
}
\altaffiliation{Joint appointment}

\date{\today}

\begin{abstract}
Non-equilibrium steady states (NESSs) generally lack thermodynamic occupation laws because finite stationary circulation and a globally exact rate-ratio field cannot coexist for the same Markov generator. 
Here we construct a sector-separated geometry that overcomes this incompatibility without arresting dissipation. Entropy-production exposure-and-separation excludes the entropy-producing state~$0$ from the conditional occupation manifold while retaining it in the dissipative full graph; physical returns $i\to0\to0^\ast$ become effectively Markovian in the strong-bias/rapid-reset (SR) limit. For a thermodynamically complete conditional manifold, autonomous redistribution (AR) eliminates residual futile circulation, making the rate-ratio one-form exact. 
Thermodynamic calibration gives $X_i=\beta(\Delta\mu- \mathcal F_i^{\mathrm{cost}})$ and $p_i=e^{X_i}/Z_\mathcal{C}$, with $Z_\mathcal{C}=1+\sum_i e^{X_i}$. Full-graph probabilities factorize exactly as $P_\alpha=(1-P_0)p_\alpha$. In the SR limit, the kinetic factor tends to unity while $p_\alpha\to e^{X_\alpha}/Z_\mathcal C$, yielding $P_\alpha\to p_\alpha$ while finite dissipation persists. 
Near AR, integrability is lost linearly in residual cycle current whereas dissipation begins quadratically. 
In the binary zero-cycle-rank limit, occupation redistributes autonomously under maintained $\Delta\mu$ bias, yielding the inverted Fermi--Dirac law, which is applied to thermal smearing in quantum-dot lasers. 
The framework provides constructive acquisition conditions and failure diagnostics for thermodynamic occupation laws in dissipative NESSs. 
\end{abstract}

\maketitle

%%%%%%%%%%%%%
\begin{figure*}[t]
\includegraphics[width=0.89\textwidth]{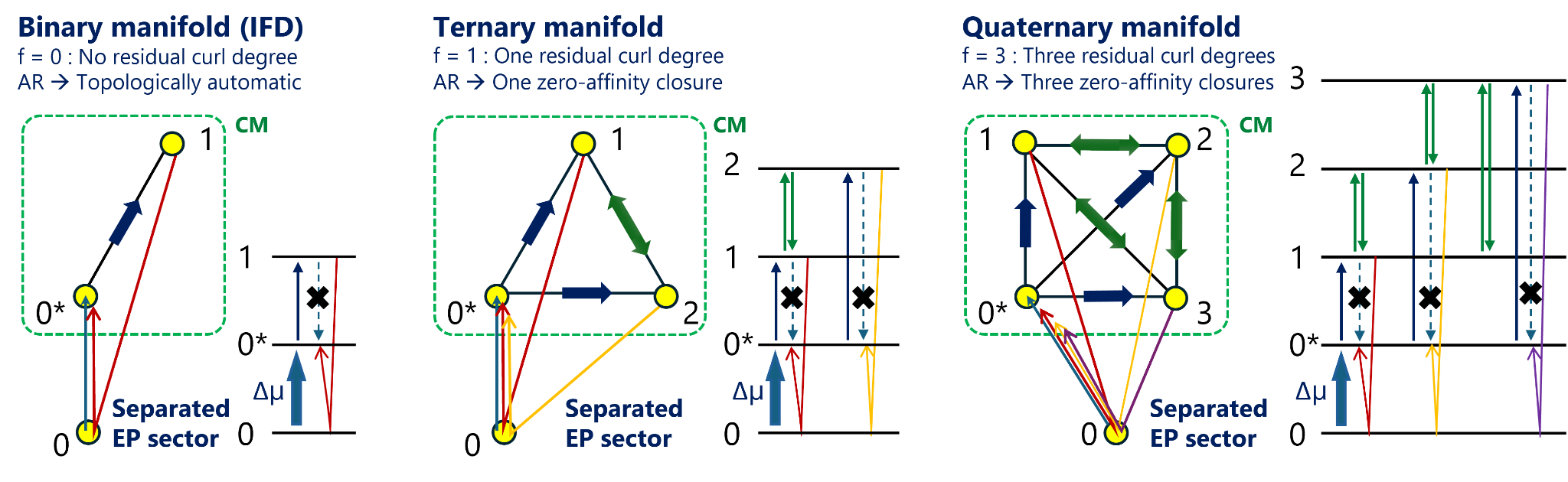}
\caption{
\textbf{EP-sector-separated dual geometry of dissipative current and thermodynamic occupation.}
Entropy-production exposure-and-separation (EPES) assigns the lower return/reset pathways to the separated entropy-producing (EP) sector.
The full graph carries the dissipative Gauss--Kirchhoff current geometry, whereas the upper nodes form the conditional manifold $\mathcal C$ (CM) of physical or mechanical elements capable of reversible energy transduction, on which the Stokes--Kolmogorov occupation geometry is defined.
Black crosses mark the prohibited direct microscopic transitions $i\to0^\ast$; instead, the physical return/reset pathway $i\to0\to0^\ast$ lies outside $\mathcal C$, passing through the separated state $0$, and its projection onto $\mathcal C$ defines the effective edge $i\to0^\ast$. 
In the rapid-reset limit, this projected edge becomes effectively Markovian. 
The binary sector-separated topology $\{0;\,0^\ast,1\}$ contains the conditional graph $\mathcal C=\{0^\ast,1\}$ with cycle rank $f=0$; no independent curl mode exists, and the theory reduces directly to the inverted Fermi--Dirac (IFD) occupation law.
The ternary topology $\{0;\,0^\ast,1,2\}$ contains $\mathcal C=\{0^\ast,1,2\}$ with $f=1$; stationarity alone leaves one residual cycle current, represented by $j_{12}=J_{12}-J_{21}$, whereas autonomous redistribution (AR) eliminates this residual mode and renders the conditional manifold curl-free.
The quaternary topology $\{0;\,0^\ast,1,2,3\}$ contains $\mathcal C=\{0^\ast,1,2,3\}$ with $f=3$ and therefore requires three independent zero-affinity closures.
In all cases, dissipation remains in the full NESS through the separated return/reset sector.
Each basin $i$ is characterized by its intrinsic state variables, the energy level $\varepsilon_i$ and degeneracy $g_i$.
}
\label{fig:sector_separated_geometry}
\end{figure*}
%%%%%%%%%%%%%

Threshold-like responses appear widely in driven non-equilibrium steady states (NESSs) such as ATP-powered skeletal muscle, semiconductor lasers, and excitable neurons
~\cite{Hill1938,Hill1989,Huxley1957,Bernard1961,Arakawa1982,HodgkinHuxley1952}. 
However, stationary occupations in a dissipative NESS are generally not thermodynamic state functions. This is because, unlike equilibrium occupations, they are entangled with probability currents, cycle affinities, traffic, and entropy production.

Coarse-graining can hide entropy-producing currents and make occupations appear equilibrium-like~\cite{Seifert2012,Esposito2012}, while hidden dissipation and violations of effective local detailed balance have been analyzed~\cite{VanderMeerErtelSeifert2022,HartichGodec2023}. 
Caliber Force Theory (CFT) provides a powerful, complete coordinate representation of a Markov NESS using node populations, edge traffic, and independent cycle fluxes~\cite{YangDill2026}. The physical foundation of the MaxCal path-entropy extremum nevertheless remains open, while recent work seeks to provide a microscopic grounding~\cite{Belousov2026}.
Lee showed that a driven Markov process violating detailed balance can be embedded in an extended Markov description that includes the driving degrees of freedom, with the original non-equilibrium process recovered after eliminating them~\cite{Lee2018}. 
Freitas and Esposito~\cite{FreitasEsposito2022} established an entropy-production bound on changes in steady-state self-information and showed that, in the macroscopic limit, saturation can occur when the coarse-grained rate-ratio field is a gradient, allowing reconstruction of the steady-state rate function.

The remaining question is how such an integrable sector can be physically realized and thermodynamically calibrated within a globally dissipative NESS. The obstruction is exposed by the discrete Helmholtz--Hodge decomposition~\cite{Polettini2015,DalCengio2023},
\begin{equation}
\mathbb{R}^{E}=\operatorname{im}\mathbf{B}^{\mathsf T} \oplus \ker \mathbf{B},
\label{eq:Hodge_decomposition}
\end{equation}
where $\mathbf B$ is the oriented incidence matrix. Gauss--Kirchhoff stationarity places $\mathbf j$ in $\ker\mathbf B$, whereas Stokes--Kolmogorov integrability places an exact one-form $\boldsymbol\omega=d\Phi$ in $\operatorname{im}\mathbf B^\mathsf T$. Although orthogonal algebraically, $\mathbf j$ and $\boldsymbol\omega$ arise from the same transition rates. On a connected irreducible Markov graph, exact $\boldsymbol\omega$ implies detailed balance and hence $\mathbf j=\mathbf 0$~\cite{Kolmogorov1936,Kelly1979}; finite current therefore requires $\boldsymbol\omega=\mathbf B^\mathsf T\Phi+\boldsymbol\omega_{\rm cyc}$. Since $\mathbf j\cdot\mathbf B^\mathsf T\Phi=0$, the entropy-production pairing is carried by $\mathbf j\cdot\boldsymbol\omega = \mathbf j\cdot\boldsymbol\omega_{\rm cyc}\neq0$. Thus, the cycle component sustaining dissipation precludes a global occupation potential.

Here, we overcome this incompatibility by reorganizing the topology (Fig.~\ref{fig:sector_separated_geometry}). Entropy-Production Exposure-and-Separation (EPES) excludes the entropy-producing state $0$ from $\mathcal C$ while retaining it in the full graph, and takes the maintained basin $0^\ast$, lifted by $\Delta\mu$, as the conditional reference. Every physical return from an active basin follows $i\to0\to0^\ast$, preserving finite dissipative throughput. Eliminating $0$ at finite reset rate generally gives semi-Markov projected dynamics. We use the strong-bias/rapid-reset (SR) regime to denote the joint hierarchy $W_{0\to0^\ast}\gg W_{\alpha\to0}$ for all $\alpha\in\mathcal C$: for $\alpha=i$ it is the rapid-reset time-scale separation that contracts $i\to0\to0^\ast$ to an effective Markov transition $i\to0^\ast$, whereas for $\alpha=0^\ast$ the same hierarchy corresponds to the strong-bias suppression of reverse pumping. Thus, the SR limit renders $\mathcal C$ effectively closed and Markovian, as required for a thermodynamic occupation description.

Another requirement is thermodynamic completeness: after EPES, the manifold $\mathcal C$ contains no unexposed non-conservative fuel input and its edges enable, in principle, reversible energy transduction. Thermodynamic completeness therefore specifies when residual conditional circulation must be interpreted as  physical futile dissipation rather than as entropy production driven by hidden fuel within $\mathcal{C}$. Once this Markovian and thermodynamically complete manifold is established, its residual irreversibility is quantified by the non-negative Schnakenberg functional. Thus, the zero-futile-dissipation reference defines autonomous redistribution (AR), for which the conditional entropy production vanishes edgewise, effective detailed balance follows, all independent cycle affinities vanish, and $\boldsymbol\omega_{\mathcal C}=d\Phi$ becomes exact, although dissipation continues in the exposed full-graph sector.

Only after exactness is established do we thermodynamically calibrate the mathematical potential. Basin-plus-environment state counting, together with reversible first-law bookkeeping and the Clausius equality, identifies $d\Phi=dX$, with $X_i=\beta(\Delta\mu-\mathcal F_i^{\rm cost})$, yielding $p_i/p_{0^\ast}=e^{X_i}$; the effective local-detailed-balance form is then recovered on $\mathcal C$. 
The cycle rank counts the independent topological obstructions; in the binary zero-cycle-rank limit, exactness is automatic and the occupation law reduces to the inverted Fermi--Dirac (IFD) form, which is applied below to a quantum-dot-laser threshold. 

Consequently, the construction separates two complementary closures. 
On the manifold $\mathcal C$, EPES, the SR Markov recovery, thermodynamic completeness and AR establish the exact Stokes--Kolmogorov occupation geometry, $p_\alpha=e^{X_\alpha}/Z_{\mathcal C}$. 
On the full graph, Gauss--Kirchhoff stationarity independently fixes the exact kinetic embedding, $P_\alpha=\Lambda_{\rm kin}p_\alpha$. In the SR limit the two factors converge separately, $\Lambda_{\rm kin}\to1$ and $p_\alpha\to e^{X_\alpha}/Z_\mathcal C$, so that $P_\alpha\to p_\alpha$ without extinguishing finite dissipative throughput. 
Upon violation of AR, residual cycle current destroys path independence at first order, whereas its energetic cost begins at second order. 
Rather than postulating a universal NESS distribution, the framework constructs a thermodynamic occupation geometry inside a dissipative current geometry, determines its kinetic embedding into the full NESS, and diagnoses when the EPES/AR construction is realized.

\bigskip
\paragraph*{\textbf{Maintenance-sector separation and conditional manifold}}

Figure~\ref{fig:sector_separated_geometry} separates the exposed entropy-producing (EP) return/reset sector from the conditional occupation manifold. The topology is written as \(\{0;\,0^\ast,1,\ldots,n\}\): the state before the semicolon belongs to the exposed sector, whereas the other nodes, after the EP sector is exposed and separated (EPES), form the conditional manifold 
\begin{equation}
\mathcal C=\{0^\ast,1,\ldots,n\},
\label{eq:ConditionalManifold}
\end{equation}
on which the occupation law is defined, where $\mathcal C$ consists of physical or mechanical elements capable of reversible energy transduction without any fuel drive.

The single-maintenance-sector rule assigns all non-conservative driving and effectively one-way, fuel-consuming chemical reactions to the exposed EP sector, including the rapid reset \(0\to0^\ast\), while every active basin returns along the serial path \(i\to0\to0^\ast\). 
Within the SR hierarchy $W_{0\to0^\ast}\gg W_{\alpha\to0}$ for all $\alpha\in\mathcal C$, the active-basin component $W_{0\to0^\ast}\gg W_{i\to0}$ makes the serial return $i\to0\to0^\ast$ representable on $\mathcal C$ by the mean-first-passage effective rate $W^{\rm eff}_{i\to0^\ast}\simeq W_{i\to0}$, while the $\alpha=0^\ast$ component suppresses reverse pumping. The entropy production associated with the reset remains assigned to the exposed EP sector; see Method~A. 
We call this construction ``thermodynamically complete'' when, at a specified maintained bias, all thermodynamic state information on $\mathcal C$ is contained in the state-dependent conservative free-energy variables of its basins, no unexposed non-conservative fuel-drive input remains within $\mathcal C$, and every edge represents a physical or mechanical element capable of reversible energy transduction.

The sector-separated topology in the full graph is normalized by
$P_0+\sum_{\alpha\in\mathcal C}P_\alpha =P_0+\Bigl(P_{0^\ast}+\sum_{i=1}^{n}P_i\Bigr)=1$.
Here \(P_0\) is the residence probability of the exposed return/reset state, and \(P_\alpha\) for \(\alpha\in\mathcal C\) are the corresponding full-topology residence probabilities. The occupation law is defined on the manifold $\mathcal{C}$, 
\begin{equation}
p_\alpha\equiv\frac{P_\alpha}{1-P_0},\quad \sum_{\alpha\in\mathcal C}p_\alpha =p_{0^\ast}+\sum_{i=1}^{n}p_i=1.
\label{eq:p_normal}
\end{equation}
Consequently, ${P_i}/{P_{0^\ast}}={p_i}/{p_{0^\ast}}$, so the common factor $1-P_0$ affects only the full-graph residence scale. Uppercase $\{P_\alpha\}$ therefore carries the Gauss--Kirchhoff flux bookkeeping, $J_{ij}=P_iW_{i\to j}$, whereas lowercase $\{p_\alpha\}$ provides the normalized probability coordinates for the Stokes--Kolmogorov occupation geometry on $\mathcal C$; 
$\{p_\alpha\}$ is determined thermodynamically, while $P_0$ is fixed by the exposed-sector kinetics.

\bigskip
\paragraph*{\textbf{Maintained bias, dissipation and thermodynamic distance}}
The maintained bias \(\Delta\mu\, (>0)\) is generated by repeated drive events on the exposed \(0\to0^\ast\) microscopic pumping edge. 
We define the one-way exposed pumping throughput as 
\begin{equation}
J_{\rm drive}\equiv J_{00^\ast}=P_0W_{0\to0^\ast}>0, 
\label{eq:JdriveJ00}
\end{equation}
where $J_{\rm drive} \,(> 0)$ is the finite turnover throughput maintaining the NESS.

In the minimal NESS model, the free-energy increment $a_\mu>0$ per drive event, the drive flux $J_{\rm drive}$, and the memory time $\tau_\mu$ define the time-averaged maintained bias
\begin{equation}
\Delta\mu\equiv\langle\chi_\mu\rangle_t
=a_\mu\tau_\mu J_{\rm drive}.
\label{eq:MaintainedBias}
\end{equation}
Thus, $\Delta\mu$ is a memory-bearing, coarse-grained bias sustained by finite throughput; its construction is given in Method~B.

Local detailed balance is first invoked at the microscopic pumping-event level; it is not imposed on the projected conditional edges. A finite reverse rate defines the affinity of one pumping event,
\begin{equation}
\ln\frac{w_{0\to0^\ast}}{w_{0^\ast\to0}} = \beta a_\mu .
\label{eq:microLDB4EP}
\end{equation}
where $w_{ij}$ denotes a microscopic event rate and \(\beta\equiv(k_{\rm B}T)^{-1}\). Here, $\beta a_\mu$ is the local total affinity of one effective pumping event, i.e., the local total entropy change of system plus reservoir in units of $k_{\rm B}$. The microscopic status of local detailed balance and its implementation in driven chemical-reaction networks are discussed in Refs.~\cite{Maes2021,RaoEsposito2016}.

The occupation distribution, however, is defined on the maintained manifold observed after coarse-graining over the memory scale $\tau_\mu$. During this memory interval, the mean number of pumping events retained in the maintained lift is $\tau_\mu J_{\rm drive}$. Accumulating the single-event affinity over this number therefore gives the $\tau_\mu$-coarse-grained pumping-edge affinity
\begin{equation}
\begin{aligned}
\ln \frac{W_{0\to0^\ast}}{W_{0^\ast\to0}}
\equiv
\tau_\mu J_{\rm drive}
\ln \Bigl(\frac{w_{0\to0^\ast}}{w_{0^\ast\to0}}\Bigr)
=\beta\Delta\mu .
\end{aligned}
\label{eq:macroLDB4EP}
\end{equation}
Equation~\eqref{eq:macroLDB4EP} fixes the $\tau_\mu$-coarse-grained forward--reverse rate ratio, while the absolute kinetic scale is fixed independently by the physical throughput $J_{\rm drive}$. 
Choosing $W_{0\to0^\ast}=J_{\rm drive}/P_0$ preserves Eq.~\eqref{eq:JdriveJ00}, and then the reverse-pump rate is given by $W_{0^\ast\to0}=W_{0\to0^\ast}e^{-\beta\Delta\mu}$. 

The entropy-production contribution assigned to the exposed drive-maintenance sector can equivalently be counted from the microscopic event frequency or from the accumulated affinity over the memory time,
\begin{equation}
\begin{aligned}
\dot S_{\rm drive}
&\simeq k_{\rm B}J_{\rm drive}
\ln\frac{w_{0\to0^\ast}}{w_{0^\ast\to0}}
=\frac{a_\mu J_{\rm drive}}{T}\\
&=\frac{k_{\rm B}}{\tau_\mu}
\ln\frac{W_{0\to0^\ast}}{W_{0^\ast\to0}}
=\frac{\Delta\mu}{\tau_\mu T}>0.
\end{aligned}
\label{eq:DriveEPbyLDB}
\end{equation}
Thus, $\ln(W_{0\to0^\ast}/W_{0^\ast\to0})=\beta\Delta\mu$ is the affinity accumulated over the memory interval, not the entropy increment of a single pumping event; division by $\tau_\mu$ converts it to an entropy-production rate. The strong-bias component of the SR approximation in Eq.~\eqref{eq:DriveEPbyLDB} uses $J_{00^\ast}-J_{0^\ast0}\simeq J_{00^\ast}=J_{\rm drive}$, with a finite microscopic reverse rate retained to define Eq.~\eqref{eq:microLDB4EP}.
Thus, the total EP rate is given by 
\begin{equation}
\dot S_{\rm tot}=\dot S_{\rm drive}+\dot {\Sigma}_{\mathcal C}. 
\label{eq:TotalEP}
\end{equation}
However, if all residual conditional modes in $\mathcal C$ are eliminated, \(\dot S_{\rm tot}=\dot S_{\rm drive}>0\). 
Hence, the full system remains a dissipative NESS even when the conditional entropy production \(\dot {\Sigma}_{\mathcal C} \,(=\dot{S}_{\mathcal C}^{\rm sys}+\dot{S}_{\mathcal C}^{\rm env})\) vanishes. 

All energetic quantities are measured using the bare reference basin~\(0\) as the energy gauge. For each active basin~\(i=1,\ldots,n\),
\begin{equation}
\Delta\varepsilon_i=\varepsilon_i-\varepsilon_0, \quad
\mathcal F_i^{\rm cost}
= \Delta\varepsilon_i - k_{\rm B}T \ln\left(\frac{g_i}{g_0}\right).
\label{eq:structural_cost}
\end{equation}
The maintained reference state \(0^\ast\) is the \(\Delta\mu\)-lifted state of the same reference basin~$0$, so that $\varepsilon_{0^\ast}-\varepsilon_0=\Delta\mu$ with \(g_{0^\ast}=g_0\). Accordingly, the signed drive--cost thermodynamic distance between the maintained reference state \(0^\ast\) and active basin~\(i\) is defined as
\begin{equation}
X_i = \beta\left( \Delta\mu-\mathcal F_i^{\rm cost} \right)
= \beta\left( a_\mu\tau_\mu J_{\rm drive}-\mathcal F_i^{\rm cost} \right).
\label{eq:ThemodynamicDistance}
\end{equation}
At this stage, $X_i$ is only a candidate thermodynamic occupation coordinate.

\bigskip
\paragraph*{\textbf{Integrability condition}}
The central question is under what conditions the scalar distance in Eq.~\eqref{eq:ThemodynamicDistance} becomes a single-valued occupation potential. 
Generally, for a connected graph on \(\mathcal{C}\) with \(V\) vertices and \(E\) undirected edges, the cycle rank is given by \(f=E-V+1\), which counts the independent internal circulation modes. By Kirchhoff--Schnakenberg network theory~\cite{Schnakenberg1976}, stationarity imposes current conservation at every node, but this alone does not eliminate the \(f\) cycle currents. 
As shown in Fig.~\ref{fig:sector_separated_geometry}, for complete binary, ternary, and quaternary conditional graphs, \(f=0,1,3\), respectively; more generally, for the complete conditional graph \(K_V\), \(f=(V-1)(V-2)/2\). 
Thus, the binary topology has no internal cycle-current mode ($f=0$), whereas higher-order complete-graph topologies necessarily retain cycle modes that cannot be eliminated by stationarity alone. 

For example, in the ternary graph, nodewise stationarity therefore constrains only current divergence, leaving one cycle mode because \(f=1\). With \(j_{12}=J_{12}-J_{21}\), stationarity of the two active basins gives
\begin{equation}
\begin{aligned}
\dot P_1&=P_{0^\ast}W_{0^\ast\to1}
-P_1W^{\rm eff}_{1\to0^\ast}-j_{12}=0,\\
\dot P_2&=P_{0^\ast}W_{0^\ast\to2}
-P_2W^{\rm eff}_{2\to0^\ast}+j_{12}=0, 
\end{aligned}
\label{eq:ternary_stationarity_main}
\end{equation}
where \(W^{\rm eff}_{i\to0^\ast}\simeq W_{i\to0}\) in the rapid-reset mean-first-passage limit (see Method~A), for which the projected return is effectively Markovian.

In the SR regime, the projected dynamics on $\mathcal C$ is effectively a continuous-time Markov jump process. Its residual stochastic irreversibility can therefore be quantified independently by the non-negative Schnakenberg production functional~\cite{Schnakenberg1976}, 
\begin{equation}
\begin{aligned}
&\dot \Sigma_{\mathcal C}
=k_{\rm B}\sum_{\{a,b\}\in E(\mathcal C)}(J_{ab}-J_{ba})\ln\frac{J_{ab}}{J_{ba}}\\
&=k_{\rm B}\sum_{\{a,b\}\in E(\mathcal C)}\left(P_aW^{(\mathrm{eff})}_{a\to b}
-P_bW^{(\mathrm{eff})}_{b\to a}\right)
\ln\frac{p_aW^{(\mathrm{eff})}_{a\to b}}
{p_bW^{(\mathrm{eff})}_{b\to a}}\ge 0, 
\end{aligned}
\label{eq:CondResidualEP}
\end{equation}
where each undirected effective edge is counted once, and $W^{(\mathrm{eff})}_{i\to j}$ indicates either $W_{i\to j}$ or $W^{\mathrm{eff}}_{i\to j}$.
Using \(J_{ab}=P_aW_{a\to b}^{(\mathrm{eff})}\) and \(P_a=(1-P_0)p_a\) gives the mixed \(P\)--\(p\) representation; the factor \(1-P_0\) cancels from the logarithmic ratio while \(P\) retains the physical flux scale. 
The stationary condition gives \(\dot{S}_{\mathcal C}^{\rm sys}=0\), and hence $\dot{\Sigma}_{\mathcal C}=\dot{S}_{\mathcal C}^{\rm env}$ in Eq.~\eqref{eq:TotalEP}. Thus, Eq.~\eqref{eq:CondResidualEP} quantifies the entropy generated by irreversible circulation within $\mathcal C$ and continuously exported to the environment. 
Within a thermodynamically complete $\mathcal C$, residual internal circulation represents futile dissipation within $\mathcal C$, rather than entropy production driven by a hidden fuel-driven reaction. Autonomous redistribution (AR) defines the ideal zero-futile-dissipation reference, for which $\dot{\Sigma}_{\mathcal C}=0$.

Every term in Eq.~\eqref{eq:CondResidualEP} satisfies $(x-y)\ln(x/y)\geq0$, therefore $\dot{\Sigma}_{\mathcal C}=0$ requires
\begin{equation}
P_a W^{(\mathrm{eff})}_{a\to b}=P_b W^{(\mathrm{eff})}_{b\to a} \quad \longleftrightarrow \quad p_a W^{(\mathrm{eff})}_{a\to b}=p_b W^{(\mathrm{eff})}_{b\to a}, 
\label{eq:EffectiveDB_Both}
\end{equation}
on every edge, which is equivalent to effective detailed balance on \(\mathcal C\). 
For each oriented effective edge \(a\to b\), the conditional rate-ratio (edge) one-form is defined as 
\begin{equation}
\omega_{ab}\equiv\ln\frac{W^{\rm (eff)}_{a\to b}}{W^{\rm (eff)}_{b\to a}}=\ln\frac{p_b}{p_a}, 
\label{eq:EdgeOneFormRatio} 
\end{equation}
where the second equality comes from Eq.~\eqref{eq:EffectiveDB_Both}. Along any independent conditional cycle \(\gamma\), the affinity $\mathcal A_\gamma$ vanishes because the probability ratios telescope: 
\begin{equation}
\mathcal A_\gamma=\oint_\gamma\omega
=\sum_{(a\to b)\in\gamma}\omega_{ab}
=\ln \Bigl(\prod_{(a\to b)\in\gamma}\frac{p_b}{p_a}\Bigr)=0,
\label{eq:ZeroCycleAffinityOneForm}
\end{equation}
with \(\gamma=1,\ldots,f\) in a chosen cycle basis. For the ternary manifold, stationarity leaves one common cycle current \(j_{0^\ast1}=j_{12}=j_{20^\ast}\equiv I_{\rm }\). When \(I_{\rm }\neq0\), it sustains a residual entropy-production rate \(k_{\rm B}I_{\rm }\mathcal A_{0^\ast12}>0\). 
Thus, AR eliminates the residual cycle modes left unconstrained by the stationary condition on $\mathcal{C}$ after EPES. 
Notably, in the light of the CFT theory~\cite{YangDill2026}, the maximum caliber applied to the thermodynamically complete $\mathcal C$ selects the same AR reference; see Method~C.

For a stationary conditional Markov network with positive rates, the Kolmogorov zero-affinity condition is equivalent to the existence of an effective detailed-balance stationary distribution~\cite{Kolmogorov1936,Kelly1979}. Hence, the conditional edge one-form is exact from Eq.~\eqref{eq:ZeroCycleAffinityOneForm}, so that a single-valued potential $\Phi_i$ exists such that
\begin{equation}
\omega_{ab}=\Phi_b-\Phi_a .
\label{eq:EdgeOneFormDifference}
\end{equation}
The exact edge one-form defines the potential at basin~$i$ by path integration along any oriented path $\Gamma_{0^\ast\to i}$ in $\mathcal C$, 
\begin{equation}
\Phi_i=\Phi_{0^\ast}+\sum_{(a\to b)\in\Gamma_{0^\ast\to i}}\omega_{ab},
\label{eq:TopologicalPath}
\end{equation}
where $(a\to b)$ denotes the elementary oriented edges traversed along the chosen path $\Gamma_{0^\ast\to i}$. Because the one-form is exact, this value is independent of the path. 
Combining Eqs.~\eqref{eq:EdgeOneFormRatio}--\eqref{eq:TopologicalPath} gives
\begin{equation}
\ln\frac{P_i}{P_{0^\ast}}=\ln\frac{p_i}{p_{0^\ast}}=\Phi_i-\Phi_{0^\ast}, 
\label{eq:ExplicitPotential}
\end{equation}
without fixing the additive gauge. 
Thus, the rate-ratio potential becomes an occupation potential only through the stationary AR balance on $\mathcal C$. 

Once AR has made the conditional edge one-form exact, thermodynamic completeness restricts its physical calibration to the pre-existing basin thermodynamic variables. Reversible first-law bookkeeping together with the Clausius equality then gives $\Phi_b-\Phi_a=X_b-X_a$ on every conditional edge, as detailed in Method~D. 
Hence, $\Phi-X$ is constant on $\mathcal C$, so that
\begin{equation}
\Phi_i=X_i+Const.,
\label{eq:PhiXcoordinateRelation}
\end{equation}
where \(Const.\) is independent of \(i\). 
This \(Const.\) produces only a common prefactor, and is absorbed exactly by normalization. Hence, only the gauge-invariant differences \(\Phi_i-\Phi_{0^\ast}=X_i-X_{0^\ast}\) enter the occupation law. 
Note that this identification does not arise by imposing LDB on the projected conditional graph.

\bigskip
\paragraph*{\textbf{Occupation laws on conditional and full graphs}}
Thus, only after integrability has been established does the thermodynamic distance become the occupation coordinate. In the SR regime, the projected return on $\mathcal C$ is quasi-Markovian (and becomes strictly Markovian in the SR limit); together with thermodynamic completeness and AR, this closes the conditional occupation problem. With $X_{0^\ast}=0$, we define the conditional AR partition function on $\mathcal C$ as
\begin{equation}
Z_{\mathcal C}
\equiv
\sum_{\alpha\in\mathcal C} e^{X_\alpha}
=
1+\sum_{i=1}^{n}
\exp\!\left[\beta
\left(\Delta\mu-\mathcal F_i^{\rm cost}\right)\right].
\label{eq:ConditionalPartitionFunction}
\end{equation}
Within this SR and thermodynamically complete AR closure, the conditional occupations converge to
\begin{equation}
p_\alpha\rightarrow\frac{e^{X_\alpha}}{Z_{\mathcal C}}, \quad p_{0^\ast}\rightarrow\frac{1}{Z_{\mathcal C}}, 
\label{eq:ConditionalAROccupation}
\end{equation}
for $\alpha\in\mathcal C$, with equality in the strict SR Markov limit. Thus, $Z_{\mathcal C}$ normalizes the conditional AR occupation geometry at the maintained bias $\Delta\mu$.

According to Eq.~\eqref{eq:p_normal}, independently of whether the projected conditional dynamics is already Markovian, the full and conditional residence probabilities obey the exact factorization
\begin{equation}
\begin{aligned}
\underbrace{P_\alpha}_{\text{full NESS probability}}
&=
\underbrace{(1-P_0)}_{\text{kinetic embedding}}
\times
\underbrace{p_\alpha}_{\text{conditional occupation}}.
\end{aligned}
\label{eq:general_full_embedding}
\end{equation}
Thus, the full-graph residence scale and the normalized conditional occupation are separated before either factor is taken to its SR limit.

%%%%%%%%%%%%%%%%%%%%%%%%%%%%%
\begin{figure*}[t]
\includegraphics[width=0.99\textwidth]{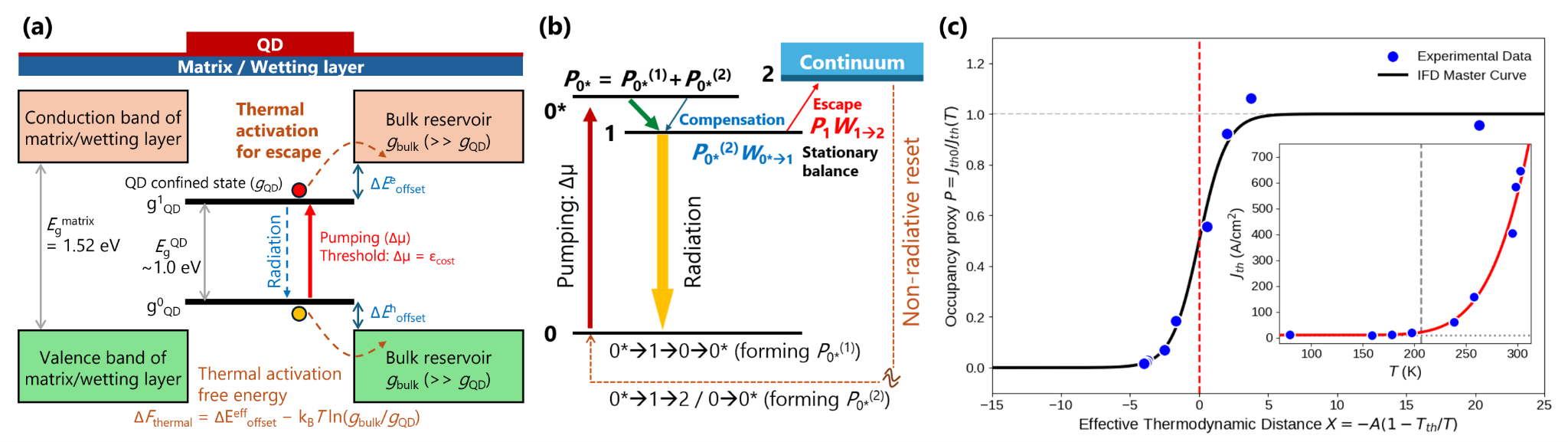}
\caption{
\textbf{Renewal-projected binary IFD mapping of a QD-laser threshold.}
\textbf{(a)} Physical source-fed topology of an InGaAs/GaAs quantum-dot (QD) laser.
Injection and pumping belong to the exposed entropy-producing sector and maintain the source state \(0^\ast\), which supplies the confined radiative QD basin~\(1\).
Once the primary Bernard--Duraffourg population-inversion condition is satisfied, thermally activated escape from the confined QD state to the high-degeneracy continuum state \(2\) provides a secondary temperature-dependent loss channel.
Direct microscopic recapture \(2\to1\) is negligible; once a carrier escapes through \(1\to2\), it is lost from the radiative branch and its loss is compensated by a new source excitation through the exposed pumping step \(0\to0^\ast\).
\textbf{(b)} Branchwise renewal bookkeeping of the threshold supply.
The maintained source-side contribution is decomposed as \(P_{0^\ast}=P_{0^\ast}^{(1)}+P_{0^\ast}^{(2)}\).
The ordinary radiative-renewal loop \(0^\ast\to1\to0\to0^\ast\) forms the baseline component \(P_{0^\ast}^{(1)}\), whereas the thermally activated escape-compensation renewal \(0^\ast\to1\to2/0\to0^\ast\) forms \(P_{0^\ast}^{(2)}\); the slash separates the escaping excitation from the newly pumped replacement and does not denote a microscopic trajectory through \(2\to0\).
At fixed temperature, the escape flux \(P_1W_{1\to2}\) is compensated by the additional source-fed supply \(P_{0^\ast}^{(2)}W_{0^\ast\to1}\).
Projection of this escape-triggered renewal onto the conditional pair \(\{0^\ast,1\}\) yields a zero-cycle-rank binary manifold with effective detailed balance.
\textbf{(c)} Digitized threshold-current data \(J_{\rm th}(T)\) from Ref.~\cite{Huffaker1998} are mapped to \(P_1=J_{\rm th0}/J_{\rm th}(T)\) and plotted against the thermodynamic coordinate \(X=\ln(R_{\rm rad}/R_{\rm esc})\).
A single global fit to all nine data points gives \(J_{\rm th0}=10.98\pm0.91~\mathrm{A\,cm^{-2}}\), \(T_{\rm th}=206.6\pm4.2~\mathrm{K}\), and \(A=12.54\pm0.55\) (\(1\sigma\)). 
The IFD-coordinate plot uses no additional fitting; it is the thermodynamic reparameterization of the same global model.
}
\label{fig:NPfig2}
\end{figure*}
%%%%%%%%%%%%%%%%%%%%%%%%%%%%%

The exposed-state probability $P_0$ is fixed independently by exact full-graph stationarity, $\dot{P_0}=0$, which requires neither the SR approximation nor Markovianity of the projected conditional dynamics; see Method E. Here $p_\alpha=P_\alpha/(1-P_0)$ is used for $\alpha\in\mathcal C$ only as the exact algebraic conditional normalization. Defining the network-return and reverse-pumping contributions by
\begin{equation}
K\equiv\sum_{i=1}^{n}p_iW_{i\to0},\quad
R\equiv p_{0^\ast}W_{0^\ast\to0},
\label{eq:ExactP0}
\end{equation}
we obtain the exact kinetic embedding factor
\begin{equation}
\begin{aligned}
\Lambda_{\rm kin}
\equiv 1-P_0
=
\left[
1+\frac{K+R}{W_{0\to0^\ast}}
\right]^{-1}.
\end{aligned}
\label{eq:KineticEmbeddingFactor}
\end{equation}
Equation~\eqref{eq:macroLDB4EP} gives $R/W_{0\to0^\ast}=p_{0^\ast}e^{-\beta\Delta\mu}$. More generally, because
$K+R=\sum_{\alpha\in\mathcal C}p_\alpha W_{\alpha\to0}$, the SR hierarchy $W_{0\to0^\ast}\gg W_{\alpha\to0}$ for all $\alpha\in\mathcal C$ directly implies $(K+R)/W_{0\to0^\ast}\ll1$ and hence $\Lambda_{\rm kin}\to1$.

For a large but finite SR separation, the pre-asymptotic product therefore remains
\begin{equation}
P_\alpha
=
\Lambda_{\rm kin}p_\alpha
\simeq
\underbrace{\left[1+\frac{K+R}{W_{0\to0^\ast}}\right]^{-1}}_{\Lambda_{\rm kin}\simeq1}
\underbrace{\frac{e^{X_\alpha}}{Z_{\mathcal C}}}_{p_\alpha\simeq e^{X_\alpha}/Z_{\mathcal C}},
\label{eq:compact_full_probability}
\end{equation}
Thus, deviations from the ideal occupation law separate into a common kinetic-residence correction and a conditional quasi-Markov correction. In the strict SR limit the two factors converge separately, 
$\Lambda_{\rm kin}\rightarrow1$,
$p_\alpha\rightarrow{e^{X_\alpha}}/{Z_{\mathcal C}}$,
so that we finally obtain
\begin{equation}
P_0\rightarrow0,
\quad
P_\alpha\rightarrow p_\alpha=\frac{e^{X_\alpha}}{Z_{\mathcal C}}.
\label{eq:FullPseudoPartition}
\end{equation}
The finite-throughput SR regime used here implies $(K+R)/W_{0\to0^\ast}\to0$ and hence $1-P_0\to1$. Thus, the exact stationary balance gives
\begin{equation}
J_{\rm drive}
=P_0W_{0\to0^\ast}
=(1-P_0)(K+R)
\rightarrow K+R.
\label{eq:RapidResetThermodynamicLimit}
\end{equation}
The strong-bias component of the same SR limit renders reverse pumping negligible, $R\to0$ (and hence $R\ll K$ for finite network-return throughput), so that $J_{\rm drive}\to K>0$. Hence, the exact pre-asymptotic factorization continuously approaches the thermodynamic Markovian conditional manifold without extinguishing the finite maintenance throughput.

Importantly, this convergence concerns the occupations on $\mathcal C$ and does not restore detailed balance on the dissipative full graph. The exposed pumping edge itself obeys the coarse-grained LDB relation in Eq.~\eqref{eq:macroLDB4EP}, whereas exact full-graph stationarity gives (see Method~E)
\begin{equation}
\frac{W_{0\to0^\ast}}{W_{0^\ast\to0}}
=e^{\beta\Delta\mu},
\quad
\frac{P_{0^\ast}}{P_0}
=
\frac{e^{\beta\Delta\mu}}{1+K/R}.
\end{equation}
Thus, whenever the network-return contribution \(K\) is finite and nonzero, the stationary residence ratio differs from the rate ratio required by detailed balance. 

The same obstruction is seen directly from Eq.~\eqref{eq:RapidResetThermodynamicLimit}: since $J_{0^\ast0}=P_{0^\ast}W_{0^\ast\to0}=(1-P_0)R$, $J_{00^\ast}-J_{0^\ast0}=P_0W_{0\to0^\ast}-P_{0^\ast}W_{0^\ast\to0}=(1-P_0)K>0$. 
Hence, away from the SR limit, the full-graph occupations generally retain a kinetic embedding correction and need not be thermodynamic-occupation-like. Only in the SR limit, where $\Lambda_{\rm kin}\to1$ and $P_\alpha\to p_\alpha$, do they inherit the Gibbs-like thermodynamic form of the conditional occupations, even though the full NESS remains dissipative.

In particular, for the binary topology $\mathcal C=\{0^\ast,1\}$, the cycle rank is $f=0$ (zero-cycle-rank).
Since $f=\dim\ker \mathbf{B}$, this is equivalent to $\ker \mathbf{B}=\{\mathbf{0}\}$.
Thus, the Helmholtz--Hodge decomposition contains no nontrivial cycle sector, so the conditional one-form is automatically exact and no additional AR cycle closure is required. 
Consequently, the binary occupation law reduces exactly to $p_1=1/(1+e^{-X_1})$ with $X_1=\beta(\Delta\mu-\mathcal F_1^{\rm cost})$. Note that the occupation itself redistributes autonomously under maintained $\Delta\mu$ bias. 
We call it the inverted Fermi--Dirac (IFD) occupation law, whose ``inversion'' expresses a subject--object reversal.
The logistic law is therefore the zero-cycle-rank corollary of the general topological theory. It is exact on the conditional pair $\{0^\ast,1\}$; the corresponding full topology $\{0;0^\ast,1\}$ approaches the same logistic probability through the exact kinetic embedding in Eq.~\eqref{eq:general_full_embedding}, with the finite-SR correction given by Eq.~\eqref{eq:compact_full_probability}.
Higher-order complete-graph manifolds, by contrast, contain genuine cycle degrees of freedom and require additional closure. 
The natural emergence of AR as the zero-cycle-rank corollary in binary systems suggests that it may extend to a more general principle of occupation redistribution in higher-order NESSs.

Here, as an experimental realization of this zero-cycle-rank corollary (IFD), we consider the thermally smeared threshold of an InGaAs/GaAs quantum-dot laser. After the primary population-inversion threshold is established, thermal escape from the confined radiative state provides a secondary loss channel. 
For this projected pair, the experimentally inferred occupation proxy $P_1=J_{\rm th0}/J_{\rm th}(T)$ plays the role of the conditional occupation, while applying Eq.~\eqref{eq:EdgeOneFormRatio} to the renewal balance gives $X=\ln(R_{\rm rad}/R_{\rm esc})$. 
Therefore, the IFD theory predicts directly $P_1=(1+e^{-X})^{-1}$.
As shown in Fig.~\ref{fig:NPfig2}, the measured threshold data follow this IFD occupation law. The full renewal topology, rate construction, and global fit are given in Methods~F and G.

\bigskip
\paragraph*{\textbf{Dissipative cost of violating the EPES/AR reference}}
\label{sec:ternary_corollary}

The EPES/AR occupation law provides a thermodynamic reference for futile dissipation. At fixed maintained bias and basin-cost landscape, AR realizes the conditional occupations without residual circulation. Because the occupation law does not fix symmetric traffic, AR defines a dissipatively optimal zero-circulation reference relative to matched-occupation states carrying residual circulation. 

To isolate this cost, we compare rate sets that preserve the AR stationary probabilities \(P_\alpha^{\rm AR}\), and hence the normalized conditional occupations \(p_\alpha^{\rm AR}\), while allowing a residual circulation. 
For example, as stated above, for the ternary cycle oriented as \(0^\ast\to1\to2\to0^\ast\), stationarity permits one divergence-free cycle-current mode, $j_{0^\ast1}=j_{12}=j_{20^\ast}\equiv I$, which preserves all nodewise stationarity conditions while breaking edgewise detailed balance as a consequence of \(I\neq0\). 

Following the traffic--cycle parametrization of Ref.~\cite{YangDill2026}, for each oriented effective edge $E$, let $q_E(I)=[J_E^+(I)+J_E^-(I)]/2$ denote the time-symmetric traffic and $\lambda_E(I)=I/[2q_E(I)]$ the dimensionless edge-current imbalance, with $q_E^{\rm AR}=q_E(0)>0$. The detailed traffic decomposition is given in Method~H. Because the occupation-ratio terms telescope around the cycle, 
\begin{equation}
\mathcal A(I)
=2\sum_E\operatorname{artanh}\lambda_E(I),
\quad
E=\{0^\ast1,12,20^\ast\}.
\label{eq:ternary_affinity_main}
\end{equation}
In the near-AR regime (not near equilibrium), the affinity expands as
\begin{equation}
\mathcal A(I)
=R_{\rm cyc}^{\rm AR}I+o(I),
\quad
R_{\rm cyc}^{\rm AR}
\equiv\sum_E\frac{1}{q_E^{\rm AR}}.
\label{eq:ternary_affinity_near_ar_main}
\end{equation}
Here, $R_{\rm cyc}^{\rm AR}$ is a kinetic calibration fixed by the balanced AR traffics, not by the occupation geometry.

When \(c=\gamma_1-\gamma_2\) is defined as the oriented cycle formed by two directed paths connecting the same pair of states on \(\mathcal C\), the discrepancy between two pathwise reconstructions of \(\Phi\) is
\begin{equation}
\delta\Phi_{\rm path}(I)
\equiv\sum_{E\in\gamma_1}\omega_E(I)-\sum_{E\in\gamma_2}\omega_E(I)
=\oint_c\omega(I)
=\mathcal A(I). 
\end{equation}
Thus, an infinitesimal residual cycle current breaks path independence at first order in $I$: the occupation potential ceases to be single-valued even when the departure from AR is arbitrarily small.
The same topological defect carries a definite energetic cost. The entropy-production rate generated within the conditional manifold is
\begin{equation}
\begin{aligned}
T \dot\Sigma_{\mathcal C}=k_{\rm B}T\,I\mathcal A(I) 
&=2k_{\rm B}T I \sum_E \operatorname{artanh}\left[\frac{I}{2q_E(I)}\right] \\
&= k_{\rm B}T R_{\rm cyc}^{\rm AR}I^2+o(I^2) \geq 0.
\end{aligned}
\label{eq:conditional_dissipation}
\end{equation}
Hence, the topological loss of integrability is linear near AR, $\delta\Phi_{\rm path}=O(I)$, whereas its energetic penalty begins at quadratic order, $T\dot\Sigma_{\mathcal C}=O(I^2)$. 
The thermodynamic state-function structure is broken before the associated dissipation becomes visible at leading linear order.

The zero-cycle-current AR reference is the matched-occupation class without futile circulation in $\mathcal{C}$. 
If~this additional dissipation is supplied through the same drive sector, its input-equivalent throughput ($\equiv \delta J_{\rm drive}$) satisfies ${a_\mu \delta J_{\rm drive}(I)}/{T} = k_{\rm B}I\mathcal A(I)$. 
At fixed maintained bias and occupation distribution, 
\begin{equation}
\Delta\mu 
= a_\mu\tau_\mu J_{\rm drive}^{\rm AR}
= a_\mu\tau_\mu\eta_{\rm alloc}(I) \left[ J_{\rm drive}^{\rm AR} + \delta J_{\rm drive}(I) \right],
\label{eq:ternary_matched_bias_main}
\end{equation}
so that we can define the allocation efficiency $\eta_{\rm alloc}$ as 
\begin{equation}
\begin{aligned}
\eta_{\rm alloc}(I)
&=
\left[
1+
\frac{k_{\rm B}T}
     {a_\mu J_{\rm drive}^{\rm AR}}
I\mathcal A(I)
\right]^{-1}
\simeq
1-
\frac{k_{\rm B}T R_{\rm cyc}^{\rm AR}}
     {a_\mu J_{\rm drive}^{\rm AR}}
I^2. 
\end{aligned}
\label{eq:ternary_allocation_efficiency_main}
\end{equation}
Thus, residual circulation generates a nonlinear affinity through \(\lambda_E=I/[2q_E(I)]\), with a linear near-AR integrability violation and a quadratic loss of input-allocation efficiency. 

Within the matched-occupation family considered here, AR therefore has a dual status. 
Mathematically, it is the integrability closure that renders the conditional occupation field exact and thereby permits thermodynamic state functions. 
Physically, it is the zero-futile-dissipation reference with a constrained variational characterization---the maximum of allocation efficiency---over stationary cycle-current deformations at the fixed occupation distribution and maintained bias.

\section*{Methods}

\subsection*{A. Effective return edges and flux bookkeeping}
An apparent conditional edge \(i\to0^\ast\) represents the physical serial return/reset pathway \(i\to0\to0^\ast\). For exponential waiting times in the minimal serial-return construction, its effective rate is defined by the mean first-passage time,
\begin{equation}
\left(W^{\rm eff}_{i\to0^\ast}\right)^{-1}
=W_{i\to0}^{-1}+W_{0\to0^\ast}^{-1}.
\label{eq:effective_return_methods}
\end{equation}
Here, \(W_{i\to0^\ast}^{\rm eff}\) is a renewal-level mean-first-passage rate rather than an instantaneous microscopic escape rate. 
In the rapid-reset limit \(W_{0\to0^\ast}\gg W_{i\to0}\), this reduces to 
\begin{equation}
W^{\rm eff}_{i\to0^\ast}\simeq W_{i\to0}. 
\label{eq:EffectiveReturnRate}
\end{equation}
The projected edge is used only on the conditional graph; the entropy-producing reset (repump) step remains assigned to the exposed sector. 

To make explicit the Markovian status of this projection, consider first a single continuous-time jump with a time-independent transition rate $W$. Let $S(t)$ denote the probability of remaining in the state up to time $t$. 
With the short-time survival probability between $t$ and $t+dt$, $S(t+dt)=[1-Wdt+o(dt)]S(t)$, the solution with $S(0)=1$ is, therefore, $S(t)=e^{-Wt}$ and the hazard function is 
\begin{equation}
\begin{aligned}
h(t)\equiv -\frac{dS/dt}{S(t)}=W.
\label{eq:single_markov_waiting}
\end{aligned}
\end{equation}
Thus, the mean waiting time is $\tau=W^{-1}$, and the constant hazard $h(t) = W$ means that the probability of a jump in the next interval $dt$ is $Wdt$, irrespective of the elapsed residence time. This is precisely the memoryless Markov property.

For the physical serial return/reset pathway $i\to0\to0^\ast$, the two successive waiting times are governed by $W_{i\to0}$ and $W_{0\to0^\ast}$. At finite reset rate, the total first-passage time is the sum of two independent exponential waiting times. 
Their convolution yields a non-exponential first-passage distribution with the time-dependent hazard rate
\begin{equation}
h_{\rm ser}(t)
=
\frac{
W_{i\to0}W_{0\to0^\ast}
\left(
e^{-W_{i\to0}t}
-
e^{-W_{0\to0^\ast}t}
\right)
}{
W_{0\to0^\ast}e^{-W_{i\to0}t}
-
W_{i\to0}e^{-W_{0\to0^\ast}t}
}.
\label{eq:serial_hazard}
\end{equation}
Accordingly, eliminating the intermediate state $0$ at finite reset time
generally produces a semi-Markov projected pathway: the completion probability
retains information about the elapsed waiting time.

In the rapid-reset limit
$W_{0\to0^\ast}\gg W_{i\to0}$, however, the residence time in the
eliminated state $0$ becomes negligible, 
\begin{equation}
\begin{aligned}
&S_{\rm ser}(t)\rightarrow e^{-W_{i\to0}t},\quad
h_{\rm ser}(t)\rightarrow W_{i\to0}.
\label{eq:rapid_reset_markov}
\end{aligned}
\end{equation}
The projected return/reset pathway therefore loses its hidden-state memory
and becomes asymptotically Markovian on $\mathcal C$. 
Consistently, the mean first-passage time reduces to $W_{i\to0}^{-1}$, so that $W^{\rm eff}_{i\to0^\ast}\simeq W_{i\to0}$, which recovers Eq.~\eqref{eq:EffectiveReturnRate}. 
Thus, the effective edge $i\to0^\ast$ is not assumed to be Markovian
after projection; the Markovianity emerges from the rapid-reset
time-scale separation. For a large but finite separation we refer to this as
the quasi-Markov rapid-reset regime, whereas the rapid-reset limit is
strictly Markovian.

Physical one-way fluxes and residual currents are evaluated on the full-topology scale,
\begin{equation}
J_{ab}=P_aW^{\rm eff}_{a\to b},\quad
j_{ab}=J_{ab}-J_{ba}.
\label{eq:flux_bookkeeping_methods}
\end{equation}
This equation is understood in the rapid-reset regime, where the reset residence is negligible and the projected renewal flux coincides with the physical flux. 
For \(a,b\in\mathcal C\), the common normalization factor cancels from the logarithmic force,
\begin{equation}
\ln\frac{J_{ab}}{J_{ba}}
=\ln\frac{P_aW^{\rm eff}_{a\to b}}{P_bW^{\rm eff}_{b\to a}}
=\ln\frac{p_aW^{\rm eff}_{a\to b}}{p_bW^{\rm eff}_{b\to a}}.
\label{eq:mixed_Pp_methods}
\end{equation}

\subsection*{B. Time-averaged construction of the maintained bias}
Let the interval between successive drive events be $\Delta t_{\rm drive}=J_{\rm drive}^{-1}$. If each event lifts the reference level by $a_\mu$ and the lift relaxes exponentially with memory time $\tau_\mu$, the maintained elevation between events is $\chi_\mu(t)=\sum_{m=0}^{\infty}a_\mu \exp[-(t+m\Delta t_{\rm drive})/\tau_\mu]$. Its average over one drive interval is 
\begin{equation}
\begin{aligned}
\langle\chi_\mu\rangle_t
= \frac{1}{\Delta t_{\rm drive}} 
\int_0^{\Delta t_{\rm drive}} \frac{a_\mu e^{-t/\tau_\mu}} {1-e^{-\Delta t_{\rm drive}/\tau_\mu}}\,dt
= a_\mu\tau_\mu J_{\rm drive}. 
\end{aligned}
\label{eq:MaintainedBiasAverageMethods}
\end{equation}
which gives Eq.~\eqref{eq:MaintainedBias}.

\subsection*{C. Maximum-caliber interpretation of AR}
For the single-cycle ternary manifold considered here, in the absence of an independent conditional cycle force, the maximum-caliber extremum has $\mathcal A=0$ and hence $I=0$~\cite{YangDill2026}; thus, within $\mathcal C$, its cycle-sector extremum is equivalent to AR. Conversely, a residual conditional cycle force generates $I\neq0$, which in the present construction signifies leakage of non-conservative driving into $\mathcal C$ and hence futile dissipation, while the exposed throughput $J_{\rm drive}>0$ remains finite. 
For higher cycle rank, vanishing all independent conditional cycle forces selects the zero-affinity AR sector.

The traffic--current parametrization used in the main text also permits a direct comparison with Caliber Force Theory. Yang and Dill denote an independent cycle flux by $J_c$; for the single-cycle ternary manifold, we identify that cycle
coordinate with the current $I$ used above. In CFT, the force conjugate to $I$ is the negative derivative of the path entropy with respect to that cycle current, with the remaining dynamical coordinates held fixed:
\begin{equation}
\mathfrak F_{\rm cyc}
=
-\left(
\frac{\partial\mathfrak s_{\rm path}}
     {\partial I}
\right)_{\{p,q_E\}}
=
\frac{1}{2}\mathcal A,
\label{eq:CFT_cycle_force}
\end{equation}
where $\mathcal A$ is the corresponding cycle affinity
\cite{YangDill2026}.

For the ternary cycle, vanishing cycle force therefore gives
\begin{equation}
\mathfrak F_{\rm cyc}=0
\longleftrightarrow
\mathcal A=0
\longleftrightarrow
I=0,
\label{eq:CFT_zero_force_DB}
\end{equation}
because zero cycle affinity imposes the Kolmogorov condition and hence detailed balance. 
Thus, when CFT is applied to the full network, vanishing all independent cycle forces restores the Markovian detailed balance. 

The present construction instead applies this maximum-caliber condition only to the thermodynamically complete conditional manifold $\mathcal C$ produced by EPES. The exposed sector is not subject to the zero-current condition and continues to carry finite dissipative throughput. By thermodynamic completeness, $\mathcal C$ contains no independent fuel drive; any residual non-conservative forcing within $\mathcal C$, if present, can only represent leakage from the exposed EP sector: 
\begin{equation}
\begin{aligned}
&J_{\rm drive}>0 \quad\text{in the full graph},\\
&\mathfrak F_{\rm cyc}^{\mathcal C}=0
\longleftrightarrow
I=0 \quad\text{in }\mathcal C .
\end{aligned}
\label{eq:CFT_AR_sector}
\end{equation}
Accordingly, CFT and AR share the same $I=0$ reference on $\mathcal C$. 
Their finite deformations need not coincide: the CFT cycle-force derivative is taken at fixed occupation and traffic, whereas the present near-AR construction fixes the occupation but allows $q_E(I)$ to vary. Nevertheless, because $q_E(I)\to q_E^{\rm AR}$ as $I\to0$, the two descriptions share the same local cycle-sector tangent and select the same zero-current extremum. 
Thus, MaxCal identifies $I=0$ as the zero-cycle-force path-entropy extremum, whereas AR independently selects the same state as the zero-futile-circulation reference of minimum residual conditional entropy production and maximum allocation efficiency. 
If non-conservative driving leaks from the exposed sector into $\mathcal C$, then $\mathfrak F_{\rm cyc}^{\mathcal C}\neq0$ generates $I\neq0$ within CFT; in the present construction, this residual circulation signals incomplete EPES and futile conditional dissipation.

\subsection*{D. Thermodynamic identification of $\Phi_i$ with $X_i$}

Once AR has rendered the conditional edge one-form exact, each node $i\in\mathcal C$ acquires a path-independent occupation potential $\Phi_i$, defined edgewise by Eq.~\eqref{eq:EdgeOneFormDifference}. At fixed $\Delta\mu$, thermodynamic completeness restricts its physical calibration to the pre-existing basin variables $\varepsilon_i$ and $g_i$ defined in Eq.~\eqref{eq:structural_cost}, i.e., $\Phi_i=\Phi(\varepsilon_i,g_i;\Delta\mu)$.

For an oriented transition $a\to b$, Eq.~\eqref{eq:structural_cost} gives
\begin{equation}
\begin{aligned}
\Delta\mathcal F_{ab}^{\rm cost}
\equiv \mathcal F_b^{\rm cost}-\mathcal F_a^{\rm cost} 
=(\varepsilon_b-\varepsilon_a)-k_{\rm B}T\ln\left(\frac{g_b}{g_a}\right),
\end{aligned}
\label{eq:BasinFreeEnergyDifference}
\end{equation}
where $\Delta S_{ab}^{\rm basin}=k_{\rm B}\ln(g_b/g_a)$.

Let $w_{ab}^{\rm rev,on}$ denote the signed reversible work done on the conditional element and $Q_{ab}^{\rm rev}$ the signed reversible heat absorbed by it. Using the first law, $\varepsilon_b-\varepsilon_a=Q_{ab}^{\rm rev}+w_{ab}^{\rm rev,on}$, together with $\Delta S_{ab}^{\rm env}=-Q_{ab}^{\rm rev}/T$, we obtain
\begin{equation}
\Delta S_{ab}^{\rm env} 
= -\frac{(\varepsilon_b-\varepsilon_a)-w_{ab}^{\rm rev,on}}{T},
\end{equation}
and hence, referring to Eq.~\eqref{eq:BasinFreeEnergyDifference},
\begin{equation}
T\left(
\Delta S_{ab}^{\rm env}
+\Delta S_{ab}^{\rm basin}
\right)
= w_{ab}^{\rm rev,on}-\Delta\mathcal F_{ab}^{\rm cost}.
\label{eq:ClausiusEqualitywithRevWork}
\end{equation}

Under AR, energy transduction along this conditional basin-to-basin isothermal transition is reversible, so the Clausius equality gives
\begin{equation}
\Delta S_{ab}^{\rm env}
+\Delta S_{ab}^{\rm basin}
=0.
\end{equation}
Therefore, Eq.~\eqref{eq:ClausiusEqualitywithRevWork} yields
\begin{equation}
w_{ab}^{\rm rev,on}
=
\Delta\mathcal F_{ab}^{\rm cost}.
\end{equation}
Thus, in this isothermal process, $\Delta\mathcal F_{ab}^{\rm cost}$ is the signed reversible free-energy cost associated with the oriented edge: it is positive when reversible work is required to move from $a$ to $b$, and negative when reversible work is released.

At this reversible AR reference, basin-plus-environment state counting based on microscopic reversibility gives the forward--reverse rate ratio associated with an oriented conditional edge as the ratio of the corresponding accessible state multiplicities, 
\begin{equation}
\begin{aligned}
\omega_{ab}
&\equiv
\ln\frac{W^{\rm (eff)}_{a\to b}}{W^{\rm (eff)}_{b\to a}}
=
\ln\frac{g_b\,\Omega_{\rm env}(E_{\rm tot}-\varepsilon_b)}
        {g_a\,\Omega_{\rm env}(E_{\rm tot}-\varepsilon_a)}
\\
&=
\ln\frac{g_b}{g_a}
+
\ln\frac{\Omega_{\rm env}(E_{\rm tot}-\varepsilon_b)}
        {\Omega_{\rm env}(E_{\rm tot}-\varepsilon_a)}
\\
&\simeq
\ln\frac{g_b}{g_a}
-
\frac{\varepsilon_b-\varepsilon_a}{k_{\rm B}T}
=
-\beta\Delta\mathcal F_{ab}^{\rm cost}.
\end{aligned}
\label{eq:state_counting_omega}
\end{equation}
Here $\Omega_{\rm env}$ denotes the number of accessible environmental microstates associated with the basin-plus-environment state-counting comparison. 
The thermodynamic-reservoir approximation in the third line follows from $S_{\rm env}=k_{\rm B}\ln\Omega_{\rm env}$ and $\partial S_{\rm env}/\partial E=1/T$. 

Because higher $\Phi_i$ corresponds to higher $p_i$, $\Phi$ has the Massieu sign convention, opposite to the usual free-energy convention. 
In the thermodynamic-reservoir limit, combining the state-counting relation in Eq.~\eqref{eq:state_counting_omega} with the reversible-work relation above calibrates the negative edge one-form by the reversible free-energy cost in units of $k_{\rm B}T$:
\begin{equation}
\begin{aligned}
-\omega_{ab}
&=
\beta w_{ab}^{\rm rev,on}
=
\beta\Delta\mathcal F_{ab}^{\rm cost}
=
X_a-X_b .
\end{aligned}
\label{eq:ThermodynamicEdgeCalibration}
\end{equation}
The last equality follows directly from Eq.~\eqref{eq:ThemodynamicDistance}. The maintained reference node is included consistently: because $\varepsilon_{0^\ast}-\varepsilon_0=\Delta\mu$ and $g_{0^\ast}=g_0$, its corresponding free-energy cost is $\mathcal F_{0^\ast}^{\rm cost}=\Delta\mu$, and hence $X_{0^\ast}=0$.

AR has already established the exact conditional one-form $\omega_{ab}=\Phi_b-\Phi_a$. Therefore, combining Eqs.~\eqref{eq:EdgeOneFormDifference} and~\eqref{eq:ThermodynamicEdgeCalibration} gives, on every conditional edge,
\begin{equation}
\Phi_b-\Phi_a
=\omega_{ab}
=X_b-X_a, 
\label{eq:ClausiusPhiXClosure}
\end{equation}
where AR supplies the integrability required for a single-valued occupation potential, whereas the basin-plus-environment state counting, together with first-law bookkeeping and the Clausius equality, supplies its physical calibration. 
Because Eq.~\eqref{eq:ClausiusPhiXClosure} holds edgewise throughout the connected manifold $\mathcal C$, $d\Phi=dX$ holds, leading to Eq.~\eqref{eq:PhiXcoordinateRelation}. 
Thus, once AR has established integrability and the resulting occupation potential has been thermodynamically calibrated, local detailed balance is recovered for the effective conditional rates,
\begin{equation}
\omega_{ab} = \ln\frac{W^{\rm (eff)}_{a\to b}}{W^{\rm (eff)}_{b\to a}} = X_b-X_a .
\end{equation}

The constant in Eq.~\eqref{eq:PhiXcoordinateRelation} produces only a common prefactor, namely, 
\begin{equation}
p_i
= 
\frac{\exp(\Phi_i)}
{\sum_{\alpha\in\mathcal C}\exp(\Phi_\alpha)}
=
\frac{\exp(X_i)}
{\sum_{\alpha\in\mathcal C}\exp(X_\alpha)},
\end{equation}
and is absorbed exactly by normalization. Hence, only the gauge-invariant differences $\Phi_i-\Phi_{0^\ast}=X_i-X_{0^\ast}$ enter the occupation law.

\subsection*{E. Kinetic determination of $P_0$ and the strong/rapid-reset limit}

The exposed-state residence probability $P_0$ is not determined by the conditional thermodynamic occupation law on $\mathcal C$. Rather, for a specified sector-separated full-graph subensemble, it is fixed by the kinetics of the exposed sector. Importantly, the stationarity relations in this subsection are full-graph identities and require neither the SR approximation nor Markovianity of the projected conditional dynamics; the normalized $p_\alpha=P_\alpha/(1-P_0)$ enter here only as algebraic conditional probabilities. The SR condition enters later, when the projected conditional manifold is made Markovian and the asymptotic embedding limit is taken.

For the minimal single-maintenance-sector topology considered here, in which the exposed state $0$ has no direct outgoing activation channel to $i\in\mathcal C$ other than the pumping step $0\to0^\ast$, exact stationarity of state $0$ gives
\begin{equation}
\dot P_0
=
\sum_{i=1}^{n}P_iW_{i\to0}
+
P_{0^\ast}W_{0^\ast\to0}
-
P_0W_{0\to0^\ast}
=0.
\end{equation}
Using $P_i=(1-P_0)p_i$ and $P_{0^\ast}=(1-P_0)p_{0^\ast}$, we define
\begin{equation}
K\equiv\sum_{i=1}^{n}p_iW_{i\to0},
\quad
R\equiv p_{0^\ast}W_{0^\ast\to0}.
\end{equation}
The exact stationary balance is therefore
\begin{equation}
P_0W_{0\to0^\ast}
=
(1-P_0)(K+R),
\label{eq:P0stationarity}
\end{equation}
which gives
\begin{equation}
P_0
=
\frac{K+R}{W_{0\to0^\ast}+K+R}.
\label{eq:P0fraction}
\end{equation}
Using the $\tau_\mu$-coarse-grained pumping-edge affinity in Eq.~\eqref{eq:macroLDB4EP}, $W_{0^\ast\to0}/W_{0\to0^\ast}=e^{-\beta\Delta\mu}$, this is equivalently
\begin{equation}
P_0
=
\frac{K/W_{0\to0^\ast}+p_{0^\ast}e^{-\beta\Delta\mu}}
{1+K/W_{0\to0^\ast}+p_{0^\ast}e^{-\beta\Delta\mu}},
\label{eq:ExactP0Methods}
\end{equation}
and $R/W_{0\to0^\ast}=p_{0^\ast}e^{-\beta\Delta\mu}$.

Dividing Eq.~\eqref{eq:P0stationarity} by $P_0$ and using $P_{0^\ast}=(1-P_0)p_{0^\ast}$, we obtain
\begin{equation}
\frac{P_{0^\ast}}{P_0}
=
\frac{p_{0^\ast}W_{0\to0^\ast}}{K+R}
=
\frac{e^{\beta\Delta\mu}}{1+K/R}.
\end{equation}
Thus, the coarse-grained pumping-edge relation fixes the rate asymmetry of the exposed $0\leftrightarrow0^\ast$ edge, but does not in general imply the corresponding detailed-balance residence relation $P_{0^\ast}/P_0=e^{\beta\Delta\mu}$. Only the special kinetic limit $K\ll R$ would recover that relation, opposite to the finite-throughput regime considered here.

The SR hierarchy can now be stated compactly by extending the return index over the whole conditional manifold,
\begin{equation}
W_{0\to0^\ast}\gg W_{\alpha\to0},
\quad \alpha\in\mathcal C.
\label{eq:SRhierarchy}
\end{equation}
For active basins $\alpha=i$, Eq.~\eqref{eq:SRhierarchy} is the rapid-reset time-scale separation that removes the hidden-state residence from $i\to0\to0^\ast$. For $\alpha=0^\ast$, together with Eq.~\eqref{eq:macroLDB4EP}, it is precisely the strong-bias condition $W_{0\to0^\ast}/W_{0^\ast\to0}=e^{\beta\Delta\mu}\gg1$, which suppresses reverse pumping. Since
\begin{equation}
K+R
=
\sum_{\alpha\in\mathcal C}p_\alpha W_{\alpha\to0},
\end{equation}
Eq.~\eqref{eq:SRhierarchy} directly implies
\begin{equation}
\frac{K+R}{W_{0\to0^\ast}}\ll1,
\quad
P_0\simeq\frac{K+R}{W_{0\to0^\ast}}\rightarrow0,
\label{eq:P0to0}
\end{equation}
while the exact embedding factor becomes
\begin{equation}
1-P_0
=
\left[1+\frac{K+R}{W_{0\to0^\ast}}\right]^{-1}
\rightarrow1.
\end{equation}
At the same time, the active-basin part of the SR hierarchy removes the semi-Markov memory of the serial return. Thermodynamic completeness and AR can then identify the conditional occupation with $p_\alpha\to e^{X_\alpha}/Z_{\mathcal C}$, giving the two-factor convergence displayed in Eqs.~\eqref{eq:compact_full_probability}--\eqref{eq:FullPseudoPartition}.

For the finite-throughput SR asymptote used here, the exact stationary balance first gives, as $1-P_0\to1$,
\begin{equation}
J_{\rm drive}
\equiv
P_0W_{0\to0^\ast}
=
(1-P_0)(K+R)
\rightarrow K+R.
\end{equation}
The strong-bias component of the same SR regime renders reverse pumping negligible, \(R\ll K\), so that \(J_{\rm drive}\simeq K>0\). Thus, $P_0\to0$ and $P_\alpha\to p_\alpha$ on the full-graph side, while the finite throughput is retained. The same SR hierarchy has two complementary consequences: it makes the conditional projection Markovian and simultaneously drives the kinetic embedding factor to unity. The full NESS therefore remains dissipative even when its conditional-manifold occupations converge to the thermodynamic law.

\subsection*{F. Effective detailed balance in a renewal-projected QD laser}

We consider the $T$-dependent threshold of an InGaAs/GaAs quantum-dot (QD) laser using the data of Huffaker \textit{et al.}~\cite{Huffaker1998}. The laser involves two distinct threshold problems. The primary threshold is the establishment of the confined carrier population required by the Bernard--Duraffourg (BD) population-inversion condition~\cite{Bernard1961}. Once this condition is satisfied, the observed $T$-dependence becomes a secondary problem: thermal escape from the confined radiative QD state into the high-degeneracy wetting-layer or matrix continuum increases the injection required to maintain lasing; see Fig.~\ref{fig:NPfig2}(a). 

The full laser is not microscopically binary. Injection and pumping maintain the source state \(0^\ast\), which feeds the confined radiative QD state~\(1\). Thermally activated escape opens an additional channel from \(1\) to the high-degeneracy continuum state~\(2\). Direct microscopic recapture from the continuum is taken to be negligible, \(W_{2\to1}\simeq0\). Once an excitation escapes through \(1\to2\), it is lost from the radiative branch; maintaining threshold then requires a new source excitation through the exposed pumping step \(0\to0^\ast\). We denote this coupled loss-and-repump renewal by \(1\to2/0\to0^\ast\). The slash is essential: it separates two successive bookkeeping events and does not represent a microscopic trajectory of the same excitation through \(2\to0\to0^\ast\).

For the ordinary primary renewal, the physical serial return/reset pathway \(1\to0\to0^\ast\) is represented on the conditional manifold by the branch-resolved effective rate \(W_{1\to0^\ast}^{\rm eff,(1)}\), i.e., the \(i=1\) specialization of the effective return edge defined in Method~A. In the rapid-reset regime,
\[
W_{1\to0^\ast}^{\rm eff,(1)}
\simeq
W_{1\to0}.
\]
The superscript \((1)\) labels this ordinary return branch, whereas \((2)\), introduced below, labels the thermally activated escape-compensation renewal.

Before branch resolution, the stationary source-fed balance may be written schematically as
\begin{equation}
P_{0^\ast}
\left(
W_{0^\ast\to1}+W_{0^\ast\to2}
\right)
=
P_1
\left(
W_{1\to0^\ast}^{\rm eff,(1)}
+
W_{1\to2}
\right).
\label{eq:QD_full_balance_methods}
\end{equation}
The initial pumping process is assumed to place the excitation pair into the confined radiative branch rather than directly into the continuum, so that \(W_{0^\ast\to2}\simeq0\). Accordingly,
\begin{equation}
P_{0^\ast}W_{0^\ast\to1}
=
P_1W_{1\to0^\ast}^{\rm eff,(1)}
+
P_1W_{1\to2}.
\label{eq:QD_full_balance_reduced_methods}
\end{equation}
To resolve the two physical loops shown in Fig.~\ref{fig:NPfig2}(b), we decompose the source-side threshold supply as
\begin{equation}
P_{0^\ast}
=
P_{0^\ast}^{(1)}
+
P_{0^\ast}^{(2)}.
\label{eq:QD_source_decomposition_methods}
\end{equation}
The component \(P_{0^\ast}^{(1)}\) belongs to the ordinary primary renewal loop
\(0^\ast\to1\to0\to0^\ast\),
whereas \(P_{0^\ast}^{(2)}\) belongs to the thermally activated escape-compensation renewal
\(0^\ast\to1\to2/0\to0^\ast\), in which escape through \(1\to2\) is compensated by a distinct repumping event \(0\to0^\ast\).
The corresponding branchwise stationary balances are
\begin{align}
P_{0^\ast}^{(1)}W_{0^\ast\to1}
&=
P_1W_{1\to0^\ast}^{\rm eff,(1)}
\,(\text{Bernard--Duraffourg}), 
\label{eq:QD_primary_branch_methods}
\\
P_{0^\ast}^{(2)}W_{0^\ast\to1}
&=
P_1W_{1\to2}
\,(\text{Thermal smearing}).
\label{eq:QD_secondary_branch_methods}
\end{align}

At the primary Bernard--Duraffourg threshold, the pumping rate is adjusted to maintain the required radiative population, so that the excitation and ordinary return/reset clocks are matched,
\begin{equation}
W_{0^\ast\to1}
\simeq
W_{1\to0^\ast}^{\rm eff,(1)}.
\end{equation}
Equation~\eqref{eq:QD_primary_branch_methods} therefore gives
\begin{equation}
P_{0^\ast}^{(1)}
=
P_1.
\label{eq:QD_primary_proxy_identity_methods}
\end{equation}
This equality is a one-to-one excitation-number bookkeeping relation, not an equality of instantaneous microscopic residence probabilities. Normalizing the total threshold supply to unity, \(P_{0^\ast}^{(1)}+P_{0^\ast}^{(2)}=1\), then gives
\begin{equation}
P_{0^\ast}^{(2)}(T)
=
1-P_1(T).
\label{eq:QD_secondary_proxy_methods}
\end{equation}
With the experimental threshold current,
\begin{equation}
P_1(T)
=
\frac{J_{\rm th0}}{J_{\rm th}(T)},
\quad
P_{0^\ast}^{(2)}(T)
=
\frac{J_{\rm th}(T)-J_{\rm th0}}
{J_{\rm th}(T)}.
\label{eq:QD_proxy_methods}
\end{equation}
Although denoted by uppercase $P$, these experimentally inferred fractions are already normalized on the projected binary pair and therefore play the role of the conditional occupations $p$ in the general theory.

For the secondary branch, the escape event \(1\to2\) and the compensating repump \(0\to0^\ast\) form the escape-triggered renewal \(1\to2/0\to0^\ast\). On the conditional binary manifold, this renewal is represented by the branch-resolved effective edge \(1\to0^\ast\), denoted by \(W_{1\to0^\ast}^{\rm eff,(2)}\). It is not a mean-first-passage trajectory of the same excitation through the continuum and reset states. When the compensating repump is rapid compared with the waiting time for thermal escape, the renewal clock is set by the escape event itself, so that
\begin{equation}
W_{1\to0^\ast}^{\rm eff,(2)}
\simeq
W_{1\to2/0\to0^\ast}
\simeq
W_{1\to2}.
\label{eq:QD_effective_escape_methods}
\end{equation}
Here, \(W_{1\to2/0\to0^\ast}\) denotes the effective transition rate of the escape-triggered replacement renewal, not a microscopic trajectory through \(2\to0\). 
Equation~\eqref{eq:QD_secondary_branch_methods} then becomes
\begin{equation}
P_{0^\ast}^{(2)}
W_{0^\ast\to1}
=
P_1
W_{1\to0^\ast}^{\rm eff,(2)}.
\label{eq:QD_effective_DB_methods}
\end{equation}
This is not microscopic detailed balance between states \(1\) and \(2\), nor is detailed balance imposed on the unreduced laser topology. Rather, Eq.~\eqref{eq:QD_effective_DB_methods} is the effective detailed-balance relation of the renewal-projected conditional manifold \(\mathcal C_{\rm QD}=\{0^\ast,1\}\). Because this graph has \(V=2\), \(E=1\), and cycle rank \(f=E-V+1=0\), it possesses no independent cycle mode. Its edge one-form is therefore automatically exact. 

For this secondary projected binary pair, the two directions of the effective edge are identified with the radiative and thermal-escape clocks,
\begin{equation}
W_{0^\ast\to1}
\simeq
R_{\rm rad}
=
\tau_{\rm rad}^{-1},
\quad
W_{1\to0^\ast}^{\rm eff,(2)}
\simeq
W_{1\to2}
=
R_{\rm esc}.
\label{eq:QD_clock_identification_methods}
\end{equation}
Thus, Eq.~\eqref{eq:QD_effective_DB_methods} gives 
\begin{equation}
\frac{P_1}{P_{0^\ast}^{(2)}}
=
\frac{R_{\rm rad}}{R_{\rm esc}} 
\label{eq:QD_rate_ratio_main}
\end{equation}
where the effective detailed balance is guaranteed. 
Since the present general theory gives
\(\omega=\ln(W_{a\to b}^{\rm eff}/W_{b\to a}^{\rm eff})=\Delta\Phi=\Delta X\), 
identifying the two effective binary transition rates with
\(R_{\rm rad}\) and \(R_{\rm esc}\) directly gives
\begin{equation}
X \equiv \ln\frac{P_1}{P_{0^\ast}^{(2)}}
=
\ln\frac{R_{\rm rad}}{R_{\rm esc}}. 
\label{eq:QD_X_main}
\end{equation}
Here $X$ is defined for the surviving radiative occupation rather than for the thermally activated escape branch. Accordingly, the natural thermal-activation coordinate has the opposite sign, $\ln(R_{\rm esc}/R_{\rm rad})=-X$: increasing temperature therefore drives $X$ downward and progressively destroys the inverted radiative occupation rather than creating it. 
Thus, binary exactness promotes the logarithmic rate ratio to the thermodynamic coordinate $X$.
For the binary conditional manifold \(\mathcal C=\{0^\ast,1\}\), the cycle rank is \(f=0\), so the conditional edge one-form is automatically exact. 

Thermal escape from the discrete QD manifold into the continuum is described by
\begin{equation}
R_{\rm esc}
=
\nu_0
\left(
\frac{g_{\rm bulk}}{g_{\rm QD}}
\right)
\exp\left(
-\beta \Delta E_{\rm offset}^{\rm eff} 
\right),
\label{eq:QD_escape_rate_main}
\end{equation}
where \(g_{\rm bulk}/g_{\rm QD}\) is the effective accessible-state multiplicity ratio sampled by the escape process.
Thus,
\begin{equation}
X(T)
=
\beta \Delta E_{\rm offset}^{\rm eff} 
-
\ln\left[
\nu_0\tau_{\rm rad}
\left(
\frac{g_{\rm bulk}}{g_{\rm QD}}
\right)
\right]
=
A\left(\frac{T_{\rm th}}{T}-1\right),
\label{eq:QD_X_T_main}
\end{equation}
with a parameter 
\(A\equiv\ln[\nu_0\tau_{\rm rad}(g_{\rm bulk}/g_{\rm QD})]\)
and the threshold temperature 
\(T_{\rm th}\equiv\Delta E_{\rm offset}^{\rm eff}/(k_{\rm B}A)\). 
The temperature-independent factor $\nu_0\tau_{\rm rad}$ provides a kinetic calibration offset in $X$, whereas $g_{\rm bulk}/g_{\rm QD}$ is the state-space multiplicity contributing to $\Delta S^{\rm eff}$. Importantly, the continuum therefore enters not only as a kinetic sink but also through its large accessible-state multiplicity.
Writing \(\Delta S^{\rm eff}=k_{\rm B}\ln(g_{\rm bulk}/g_{\rm QD})\), the relevant thermal cost is the free-energy barrier
\begin{equation}
\Delta F_{\rm thermal}(T)
=
\Delta E_{\rm offset}^{\rm eff}
-
T\Delta S^{\rm eff},
\label{eq:QD_free_energy_main}
\end{equation}
rather than a purely energetic offset. Thus, Eq.~\eqref{eq:QD_X_T_main} can be written as
$
X(T)
=
\beta \left[
\Delta F_{\rm thermal}
-
k_{\rm B}T
\ln\left(
\nu_0\tau_{\rm rad}
\right)
\right]. 
$
The entropy gained by accessing the vast continuum lowers the effective thermal barrier and shifts the onset of threshold degradation without introducing an empirical temperature offset. 

Because the projected manifold is binary and cycle-free, Eq.~\eqref{eq:QD_X_main} enters directly into the IFD occupation law, giving
\begin{equation}
J_{\rm th}(T)
=
J_{\rm th0}
\left\{
1+\exp[-X(T)]
\right\}.
\label{eq:QD_IFD_main}
\end{equation}
A single global fit to all nine digitized threshold-current values is shown in Fig.~\ref{fig:NPfig2}(c). Thus, the threshold data are not merely fitted by a logistic curve: the IFD coordinate resolves the thermal degradation into an energetic offset and an experimentally accessible entropic contribution from the continuum state space.

\subsection*{G. Data collapse by IFD fitting}

All nine digitized threshold-current values were fitted globally in logarithmic current space by minimizing
\begin{equation}
\mathrm{RSS}_{\ln J}
=
\sum_k
\left[
\ln J_{\rm th}^{\rm model}(T_k)
-
\ln J_{\rm th}^{\rm data}(T_k)
\right]^2. 
\label{eq:QD_fit_objective}
\end{equation}
The low-temperature plateau \(J_{\rm th0}\), crossover temperature \(T_{\rm th}\), and dimensionless parameter \(A\) were treated as global fit parameters.
Quoted uncertainties are \(1\sigma\) parameter errors obtained from the least-squares covariance using \(N-3\) degrees of freedom.
Small excursions of \(P_1\) above unity near the plateau were retained as experimental or digitization scatter and were not interpreted as physical occupations. 
At the optimum, \(\mathrm{RSS}_{\ln J}=0.10445\) and the logarithmic root-mean-square residual is \(0.10773\).

The fit gives
\(J_{\rm th0}=10.98~\mathrm{A\,cm^{-2}}\), \(T_{\rm th}=206.6~\mathrm{K}\), and \(A=12.54\).
The corresponding effective energy offset is
\(\Delta E_{\rm offset}^{\rm eff}
=
k_{\rm B}AT_{\rm th}
=
223.2~{\rm meV}\).
Using \(\nu_0=10^{13}~\mathrm{s^{-1}}\) and \(\tau_{\rm rad}=10^{-9}~\mathrm{s}\), consistent with established QD and semiconductor-laser time scales~\cite{Zhang2000,Jiang2008,Coldren2012}, we obtain 
\(\Delta S^{\rm eff}/k_{\rm B}=3.33\),
corresponding to an effective continuum multiplicity ratio of approximately \(28\). 
The entropic contribution is also consistent with the observed onset of threshold degradation. 
Above approximately \(200\)--\(220~\mathrm{K}\), the measured threshold exhibits an Arrhenius-like rise with an activation energy of approximately \(235~{\rm meV}\). 

Finally, the conventional characteristic-temperature form~\cite{Coldren2012}, \(J_{\rm th}(T)=J_0\exp(T/T_0)\), is recovered as a local high-\(T\) limit of the IFD law.

\subsection*{H. Near-AR traffic parametrization}

For the matched-occupation ternary deformation introduced above, consider an oriented effective edge $E:i\to j$. The one-way fluxes are defined as \(J_E^+(I)\equiv P_i^{\rm AR}W_{i\to j}^{(\rm eff)}(I)\) and \(J_E^-(I)\equiv P_j^{\rm AR}W_{j\to i}^{(\rm eff)}(I)\). 
Applying the elementary identity $\ln(x/y)=2\,\operatorname{artanh}[(x-y)/(x+y)]$ to Eq.~\eqref{eq:CondResidualEP} the logarithmic edge force is naturally expressed in terms of the difference and sum of the two one-way fluxes. 
Then, $I$ is expressed self-consistently as 
\begin{equation}
I = J_E^+(I)-J_E^-(I), 
\end{equation} 
and  $q_E(I)$ is defined as
\begin{equation}
\begin{aligned}
2q_E(I) \equiv J_E^+(I)+J_E^-(I). 
\end{aligned}
\label{eq:ternary_traffic_coordinates_main}
\end{equation}
Here, \(2q_E\) is the time-symmetric edge traffic and \(I\) the net cycle current; this sum--difference parametrization is algebraically equivalent to the traffic--cycle-flux coordinates used in Caliber Force Theory~\cite{YangDill2026}. 

The dimensionless edge-current imbalance is then 
\begin{equation}
\frac{J_E^+(I)-J_E^-(I)}{J_E^+(I)+J_E^-(I)} =\frac{I}{2q_E(I)}\equiv \lambda_E(I)  ,
\label{eq:ternary_edge_imbalance_main}
\end{equation}
where $|\lambda_E(I)|<1$. 
Thus, $J_E^\pm(I) = q_E(I)\left[1\pm\lambda_E(I)\right]$, identically. 
This decomposition imposes no restriction on the variation of \(q_E(I)\).
At the EPES/AR reference, \(\lambda_E(0)=0\), and effective detailed
balance gives
\begin{equation}
q_E^{\rm AR}\equiv q_E(0) = P_i^{\rm AR}W_{i\to j}^{\rm (eff)}(0) = P_j^{\rm AR}W_{j\to i}^{\rm (eff)}(0).
\label{eq:ternary_ar_balanced_traffic_main}
\end{equation}
Different values of \(q_E^{\rm AR}\) represent different balanced kinetics realizing the same AR occupations.

The corresponding conditional edge one-form is
\begin{equation}
\begin{aligned}
&\omega_E(I)
= \ln \frac{W_{i\to j}^{\rm (eff)}(I)}{W_{j\to i}^{\rm (eff)}(I)} = \ln \frac{J_E^+(I)/P_i^{\rm AR}}{J_E^-(I)/P_j^{\rm AR}}
\\
&\quad= \ln\frac{p_j^{\rm AR}}{p_i^{\rm AR}} + \ln\frac{1+\lambda_E(I)}{1-\lambda_E(I)}
= \ln\frac{p_j^{\rm AR}}{p_i^{\rm AR}} + 2\operatorname{artanh}\lambda_E(I).
\end{aligned}
\label{eq:ternary_edge_oneform_main}
\end{equation}
Because the occupation-ratio terms telescope around the closed cycle, the cycle affinity becomes Eq.~\eqref{eq:ternary_affinity_main}. 
Thus, at matched occupation (i.e., with $p_i$ and $\Delta \mu$ both invariant), the violation of Kolmogorov closure is governed by the relative edge imbalances \(\lambda_E=I/(2q_E)\). 
In the near-AR regime, \(|\lambda_E|\ll 1\), if \(q_E(I)\to q_E^{\rm AR} \,(>0)\) as \(I\to0\), then Eq.~\eqref{eq:ternary_affinity_near_ar_main} is obtained.

\subsection*{AI-assisted editing}
During the preparation of this work, the author used large language models for English-language editing and for adversarial critical-reading stress tests intended to identify possible misinterpretations of the manuscript. All scientific reasoning, verification, interpretation and final judgements were performed by the author, who takes full responsibility for the scientific content.

\section*{Data Availability}
The digitized dataset used in Fig.~\ref{fig:NPfig2} is available from the author upon reasonable request. 

\section*{Code Availability}
No specialized software or custom simulation code was used. Any scripts used only for digitizing, fitting and plotting the data in Fig.~\ref{fig:NPfig2} are available from the author upon reasonable request.

\bibliographystyle{apsrev4-2}
\bibliography{References_checked}

\section*{Acknowledgements}
The author notes that the conception of this study dates back to discussions during his student days, and wishes to express his gratitude to Professor Katsushi Tanaka (currently at Kobe University, formerly at Kyoto University) for the discussions at that time.

\section*{Author Contributions}
T.I. solely conceived the study, developed the theoretical framework, and wrote the manuscript.

\section*{Competing Interests}
The author declares no competing interests.

\end{document}